# Quantum Query Algorithms for Certain Problems and Generalization of Algorithm Designing Techniques

Alina Dubrovska, Taisia Mischenko-Slatenkova
Department of Computer Science, University of Latvia,
Raina bulv. 29, LV-1459, Riga, Latvia
Alina.Dubrovska@gmail.com, TaisiaMischenko@gmail.com

**Abstract.** Quantum algorithms can be analyzed in a query model to compute Boolean functions where input is given in a black box, but the aim is to compute function value for arbitrary input using as few queries as possible. In this paper we concentrate on quantum query algorithm designing tasks. The main aim of research was to find new efficient algorithms and develop general algorithm designing techniques. We present several exact quantum query algorithms for certain problems that are better than classical counterparts. Next we introduce algorithm transformation methods that allow significant enlarging of sets of exactly computable functions. Finally, we propose algorithm constructing methods applicable for algorithms with specific properties that allow constructing algorithms for more complex functions preserving acceptable error probability and number of queries.

## 1 Introduction

Let $f(x_1, x_2, ..., x_n) : \{0,1\}^n \to \{0,1\}$ be a Boolean function. We study the query model, where the input $(x_1, x_2, ..., x_n)$ is contained in a black box and can be accessed by asking questions about the values of $x_i$. The goal here is to compute the value of function. The complexity of a query algorithm is measured in number of questions that it asks. The classical version of this model is known as *decision trees* [1]. Quantum query algorithms can solve certain problems faster than classical algorithms and best-known exact quantum algorithm was designed for PARITY function with $n/2$ questions vs. $n$ questions required by classical algorithm [2,3].

The problem of quantum algorithm construction, exact in particular, is not rather easy. Although there are numerous quantum algorithm complexity lower and upper bound estimations [2] examples of non-trivial and original quantum query algorithms are very few. Moreover, there is no special technique described to build a quantum algorithm for a certain function with defined in advance complexity and number of variables.

Boolean functions are widely adopted in real life processes, that is why our capacity to build a quantum algorithm for an arbitrary function appears to be extremely important. While working on common techniques, we try to collect examples of efficient quantum algorithms to build up a base for powerful computation using the advantages of quantum computer.

# 2 Notation and Definitions

Let $f(x_1, x_2, ..., x_n): \{0,1\}^n \to \{0,1\}$ be a Boolean function. We use $\oplus$ to denote XOR operation (exclusive OR). We use $\bar{f}$ for the function $1 - f$. We also use abbreviation QQA for "quantum query algorithm".

## 2.1 Quantum computing

We apply the basic model of quantum computing. For more details, see textbooks by Gruska [4] and Nielsen and Chuang [5].

An $n$-dimensional quantum pure state is a vector $|\psi\rangle \in C^n$ of norm 1. Let $|0\rangle, |1\rangle, ..., |n-1\rangle$ be an orthonormal basis for $C^n$. Then, any state can be expressed as $|\psi\rangle = \sum_{i=0}^{n-1} a_i |i\rangle$ for some $a_i \in C^n$. Since the norm of $|\psi\rangle$ is 1, we have $\sum_{i=0}^{n-1} |a_i|^2 = 1$. States $|0\rangle, |1\rangle, ..., |n-1\rangle$ are called *basic states*. Any state of the form $\sum_{i=0}^{n-1} a_i |i\rangle$ is called a *superposition* of $|0\rangle, |1\rangle, ..., |n-1\rangle$. The coefficient $a_i$ is called an *amplitude* of $|i\rangle$.

The state of a system can be changed using *unitary transformations*. Unitary transformation $U$ is a linear transformation on $C^n$ that maps vector of unit norm to vectors of unit norm. If, before applying $U$, the system was in state $|\psi\rangle$, then the state after the transformation is $U|\psi\rangle$.

The simplest case of quantum measurement is used in our model. It is the full measurement in the computation basis. Performing this measurement on state $|\psi\rangle = a_1|0\rangle + ... a_k|k\rangle$ gives the outcome $i$ with probability $|a_i|^2$. The measurement changes the state of the system to $|i\rangle$ and destroys the original state $|\psi\rangle$.

## 2.2 Query model

Query model is probably the simplest model for computing Boolean functions. In this model, the input $x_1, x_2, ..., x_n$ is contained in a black box and can be accessed by asking questions about the values of $x_i$. Query algorithm must be able to determine the value of a function correctly for arbitrary input contained in a black box. The complexity of the algorithm is measured by the number of queries to the black box that it uses. The classical version of this model is known as *decision trees*. For details, see the survey by Buhrman and de Wolf [1].

We consider computing Boolean functions in the quantum query model. For more details, see the survey by Ambainis [6] and textbooks by Gruska [4] and de Wolf [2]. A quantum computation with $T$ queries is a sequence of unitary transformations:

$$U_0 \to Q_0 \to U_1 \to Q_1 \to ... \to U_T \to Q_{T-1} \to U_T$$

$U_i$'s can be arbitrary unitary transformations that do not depend on the input bits $x_1, x_2, ..., x_n$.

$Q_i$'s are query transformations. Computation starts in a state $|\vec{0}\rangle$. Then we apply

$U_0, Q_0, ..., Q_{T-1}, U_T$ and measure the final state.

There are several different, but equally acceptable ways to define quantum query algorithms. The most important consideration is to choose an appropriate definition for the query black box, defining way of asking questions and receiving answers from the oracle.

Next we will precisely describe the full process of quantum query algorithm definition and notation used in this paper.

Each quantum query algorithm is characterized by the following parameters:

1) *Unitary transformations*

All unitary transformations and the sequence of their application (including the query transformation parts) should be specified. Each unitary transformation is a unitary matrix.

Here is an example of an algorithm sequence specification with *T* queries:

$$|\vec{0}\rangle \to U_0 \to Q_1 \to ... \to Q_T \to U_N \to [QM],$$

where $|\vec{0}\rangle$ is initial state, [QM] – quantum measurement.

For convenience we will use *bra* notation for describing state vectors and algorithm flows. Quantum mechanics employs the following notation for state vectors [5]:

$$\text{Ket notation:} |\psi\rangle = \begin{pmatrix} \alpha_1 \\ ... \\ \alpha_n \end{pmatrix} \qquad \text{Bra notation:} \langle\psi| = |\psi\rangle^+ = (\alpha_1^*, ..., \alpha_n^*)$$

Algorithm designed in *bra* notation can be converted to *ket* notation by replacing each unitary transformation matrix with its adjoint matrix (conjugate transpose):

Quantum query algorithm flow in *bra* notation: $\langle\psi| = \langle\vec{0}| U_0 Q_0 ... Q_{N-1} U_N$

Quantum query algorithm flow in *ket* notation: $|\psi\rangle = U_N^+ Q_{N-1}^+ ... Q_0^+ U_0^+ |\vec{0}\rangle$

2) *Queries*

We use the following definition of query transformation - if input is a state $|\psi\rangle = \sum_i a_i |i\rangle$, then the output is $|\phi\rangle = \sum_i (-1)^{x_k} a_i |i\rangle$, where we can arbitrary choose variable assignment $x_k$ for each amplitude. Assume we have a quantum state with *m* amplitudes $\langle\psi| = (\alpha_1, \alpha_2, ..., \alpha_m)$. For the *n* argument function, we define a query as $QQ_i = (\alpha_1 \equiv k_1, ..., \alpha_m \equiv k_m)$, where *i* is the number of question, and $k_j \in \{1..n\}$ is the number of queried variable (*QQ* abbreviates "quantum query"). If $x_{k_j} = 1$, a query will change the sign of the *j*-th amplitude to the opposite sign; in other case, the sign will remain as-is. Unitary matrix that corresponds to query transformation $QQ_i = (\alpha_1 \equiv k_1, ..., \alpha_m \equiv k_m)$ is:

$$QQ = \begin{pmatrix} (-1)^{k1} & 0 & ... & 0 \\ 0 & (-1)^{k2} & ... & 0 \\ ... & ... & ... & ... \\ 0 & 0 & ... & (-1)^{km} \end{pmatrix}$$

3) *Measurement*

Each amplitude of a final quantum state corresponds to the algorithm output. We assign a value of a function to each output. We denote it as $QM = (\alpha_1 \equiv k_1, ..., \alpha_m \equiv k_m)$, where $k_i \in \{0,1\}$ (*QM* abbreviates "quantum measurement"). The result of running algorithm on input *X* is *j* with a probability that equals the sum of squares of all amplitudes, which corresponds to outputs with value *j*. The following diagram represents the query algorithm in general form:

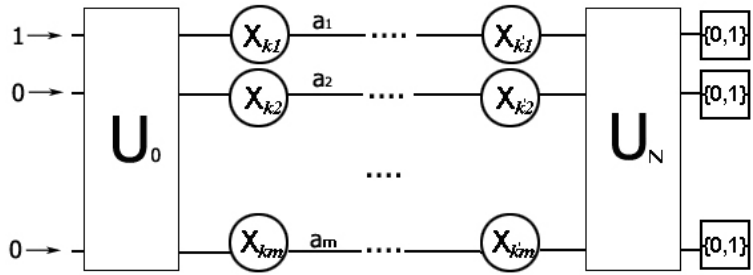

Fig. 1 Graphical representation of a quantum query algorithm.

## 2.3 Query Algorithm Complexity

The complexity of a query algorithm is based on the number of questions it uses to determine the value of a function on worst-case input. All the definitions below are adopted from [1].

The *deterministic complexity* of a function *f*, denoted by *D(f)*, is the minimum number of queries that must be asked on any input by an optimal deterministic algorithm for *f*.

For deterministic query complexity estimation for a function, the notion of sensitivity *s(f)* is useful. The sensitivity of *f* on input $(x_1, x_2, \ldots, x_n)$ is the number of variables $x_i$ with the following property: $f(x_1,\ldots,x_i,\ldots,x_n) \neq f(x_1,\ldots,1-x_i,\ldots,x_n)$. The sensitivity of *f* is the maximum sensitivity of all possible inputs. It has been proved that $D(f) \geq s(f)$.

A quantum query algorithm *computes f exactly* if the output equals *f(x)* with a probability 1, for all $x \in \{0,1\}^n$. $Q_E(f)$ denotes the number of queries of an optimal exact quantum query algorithm for a function *f*.

A quantum query algorithm *computes f with bounded-error* if the output equals *f(x)* with probability at least 1/2, for all $x \in \{0,1\}^n$. $Q_P(f)$ denotes the number of queries of an bounded-error quantum query algorithm for a function *f* that produces correct answer with probability *p*.

## 3 Exact Quantum Query Algorithms for Certain Problems

In this section we present exact quantum query algorithms for several certain problems. We start with relatively simple problems and reduce them to the task of computing Boolean functions. We design quantum query algorithms that compute target Boolean functions without error probability and are better than best possible classical query algorithms. We would like to emphasize that for a moment the best known separation between classical deterministic and exact quantum algorithm complexity is *n* vs. $\left\lceil \frac{n}{2} \right\rceil$ for PARITY function [2,3]. In our algorithms we do not exceed this limit and do not show the new best record, but our algorithms obtain exactly the same complexity gap.

### 3.1 3-variable function with 2 queries

In this section we present quantum query algorithm for 3-variable Boolean function that saves one query comparing to the best possible classical deterministic algorithm.

**Problem 1:** *Check if all input variable values are equal.*

Possible real life application is, for example, automated voting system, where statement is automatically approved only if all participants voted for acceptance/rejection equally. We provide solution for 3-party voting routine. Our algorithm needs only 2 queries, though any classical algorithm would require all 3 queries. We reduce Problem 1 to computing the following Boolean function defined by the logical formula:

$$EQUALITY_3(X) = \neg(x_1 \oplus x_2) \wedge \neg(x_2 \oplus x_3)$$

It is easy to see, that $EQUALITY_3(X)=1$ iff all variable values are equal.

**Deterministic complexity:** $D(EQUALITY_3)=3$. Provided by sensitivity on any accepting input.

**Algorithm 1.** Exact quantum query algorithm *for EQUALITY₃* is presented in Figure 2. Each horizontal line corresponds to the amplitude of basic state. Computation starts with amplitude distribution $\langle \vec{0} | = (1,0,0,0)$. Three large rectangles correspond to 4x4 unitary matrices. Two vertical layers of circles specify queried variable order for each query. Finally, four small squares at the end of each horizontal line define assigned function value for each output.

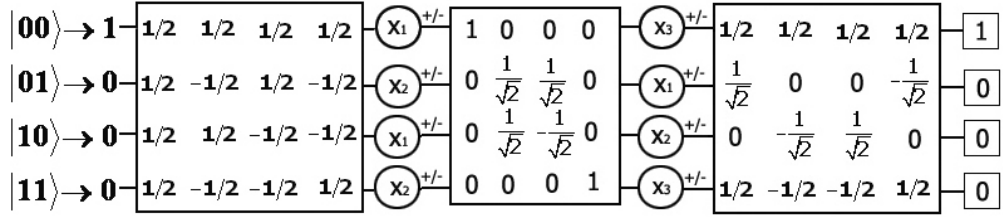

Fig. 2 Exact quantum query algorithm for *EQUALITY*.

To demonstrate quantum query algorithm processing flow for inexperienced reader we show computations for accepting and rejecting inputs.

| Input | Computation flow | Result |
|---|---|---|
| 011 | $\langle \psi | = \left(\frac{1}{2},\frac{1}{2},\frac{1}{2},\frac{1}{2}\right) Q_0 U_1 Q_1 U_2 = \left(\frac{1}{2},-\frac{1}{2},\frac{1}{2},-\frac{1}{2}\right) U_1 Q_1 U_2 =$ <br> $= \left(\frac{1}{2},0,-\frac{1}{\sqrt{2}},-\frac{1}{2}\right) Q_1 U_2 = = \left(-\frac{1}{2},0,\frac{1}{\sqrt{2}},\frac{1}{2}\right) U_2 = (0,-1,0,0)$ | REJECT |
| 111 | $\langle \psi | = \left(\frac{1}{2},\frac{1}{2},\frac{1}{2},\frac{1}{2}\right) Q_0 U_1 Q_1 U_2 = \left(-\frac{1}{2},-\frac{1}{2},-\frac{1}{2},-\frac{1}{2}\right) U_1 Q_1 U_2 =$ <br> $= \left(-\frac{1}{2},-\frac{1}{\sqrt{2}},0,-\frac{1}{2}\right) Q_1 U_2 = \left(\frac{1}{2},0,\frac{1}{\sqrt{2}},\frac{1}{2}\right) U_2 = (1,0,0,0)$ | ACCEPT |

Table 1. Computation flows for Algorithm 1.

## 3.2 4-variable function with 2 queries

In this section we present our solution for well known computational problem of comparing two binary strings.

**Problem 2:** *Check if two binary strings are equal.*

We present an algorithm for strings of length 2. We reduce Problem 2 to computing the Boolean function of 4 variables. There first two variables represent the first string, but second two variables correspond to the second string that we would like to compare with the first one. Target Boolean function can be represented by logical formula:

$$STRING\_EQ_4(X) = \neg((x_1 \oplus x_3) \vee (x_2 \oplus x_4))$$

It is easy to check that $STRING\_EQ_4(X) = 1$ iff two binary strings $x_1x_2$ and $x_3x_4$ are equal.

**Deterministic complexity**: $D(STRING\_EQ_4)=4$. Provided by sensitivity on any accepting input.

**Algorithm 2**. Exact quantum query algorithm for $STRING\_EQ_4$ is presented in Figure 3.

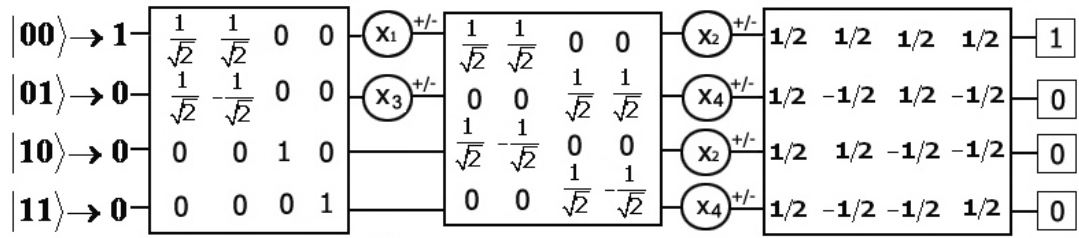

Fig. 3 Exact quantum query algorithm for STRING_EQ.

## 3.3  2*n*-variable functions with *n* queries

In this section we present a set of exact quantum query algorithms, which perform the same advantage of quantum query complexity over deterministic one as PARITY function does.
For the convenience of notation let us split the input into two parts X and Y respectively.

**Problem 3:** *Let us consider a function $T_4$ of 4 variables defined by a truth-table:*

| x1 | x2 | y1 | y2 | $T_4$(XY) |
|---|---|---|---|---|
| 0 | 0 | 0 | 0 | 1 |
| 0 | 0 | 1 | 1 | 1 |
| 0 | 1 | 0 | 1 | 1 |
| 0 | 1 | 1 | 0 | 1 |
| 1 | 0 | 0 | 1 | 1 |
| 1 | 0 | 1 | 0 | 1 |
| 1 | 1 | 0 | 0 | 1 |
| 1 | 1 | 1 | 1 | 1 |
| otherwise | | | | 0 |

It can be described as: 1 if the number of "1" in the input is even and first half of the input is either symmetric to the second half or asymmetric. We denote *Hamming weight* of the input XY=<x1,x2,y1,y2> as |XY|. Then

$T_4(x1, x2, y1, y2) = 1 \Leftrightarrow ((x1 = y1\ and\ x2 = y2) or (x1 \neq y1\ and\ x2 \neq y2))\ and\ |XY|\ is\ even$

**Deterministic complexity:** $D(T_4) = 4$. Provided by sensitivity on any input string such that $T_4$(XY) = 1 (XY = 0000, XY = 0011).

**Algorithm 3.** Exact quantum query algorithm for $T_4$ with 2 queries is presented in Figure 4.

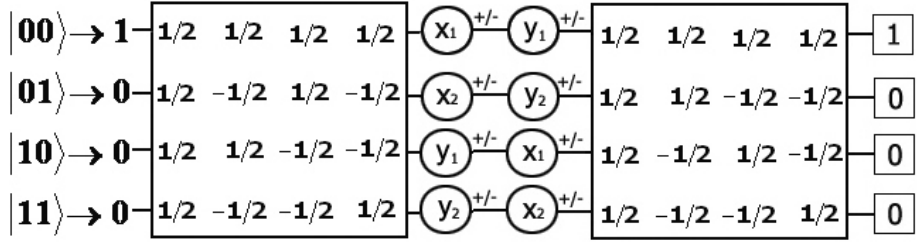

Fig. 4. Exact quantum query algorithm for $T_4$.

Computation process is specified by following sequence:

$$|0\rangle \to U_0 \to Q_1 Q_2 \to U_1 \to [QM]$$

It appears that the idea of this algorithm can be used for a bigger number of variables. For example, let us define a function $T_6$ of 6 variables:

| XY | $T_6$ | XY | $T_6$ | XY | $T_6$ | XY | $T_6$ |
|---|---|---|---|---|---|---|---|
| 000000 | 1 | 010010 | 1 | 100001 | 1 | 110011 | 1 |
| 000101 | 1 | 010111 | 1 | 100100 | 1 | 110110 | 1 |
| 001001 | 1 | 011011 | 1 | 101000 | 1 | 111010 | 1 |
| 001100 | 1 | 011110 | 1 | 101101 | 1 | 111111 | 1 |
| otherwise | 0 | | | | | | |

Which computation process is specified by following sequence:

$$|0\rangle \to U_0 \to Q_1 Q_2 Q_3 \to U_1 \to [QM], \text{ where}$$

$Q_1 = (\alpha_1 = x_1, \alpha_2 = x_2, \alpha_2 = y_1, \alpha_4 = y_2),$

$Q_2 = (\alpha_1 = x_2, \alpha_2 = x_3, \alpha_2 = y_2, \alpha_4 = y_3),$

$Q_3 = (\alpha_1 = y_1, \alpha_2 = y_3, \alpha_2 = x_1, \alpha_4 = x_3)$

In other words, value of amplitude $a_0$ depends on sequence of values of <x1,x2,y1>, $a_1$, $a_2$, $a_3$ respectively on <x2,x3,y3>, <y1,y2,x1>, <y2,y3,x3>.

It is possible to generalize this idea.

**Theorem 1.** For Boolean function $T_{2n}$ of $2n$ variables, there exists an exact quantum query algorithm which computes this function with $n$ queries.

**Proof**.

The idea of the algorithm remains the same as in the case of $T_4$ and $T_6$, we use a chain of transformations $|0\rangle \to U_0 \to Q \to U_1 \to [QM]$, where Q is a sequence of queries. Here it is more convenient to visualize Q in the form of a matrix, where $i$-th query $Q_i$ is represented by the $i$-th column of matrix Q.

For example, Q for $T_6$ is $Q = \begin{pmatrix} x1 & x2 & y1 \\ x2 & x3 & y3 \\ y1 & y2 & x1 \\ y2 & y3 & x3 \end{pmatrix}$ .

Generalizing Q for $T_{2n}$:

for even number $n$ $Q_{even} = \begin{pmatrix} x_1 & \ldots & x_{\frac{n}{2}} & y_1 & \ldots & y_{\frac{n}{2}} \\ x_{\frac{n}{2}+1} & \ldots & x_n & y_{\frac{n}{2}+1} & \ldots & y_n \\ y_1 & \ldots & y_{\frac{n}{2}} & x_1 & \ldots & x_{\frac{n}{2}} \\ y_{\frac{n}{2}+1} & \ldots & y_n & x_{\frac{n}{2}+1} & \ldots & x_n \end{pmatrix}$,

for odd $n$ $Q_{odd} = \begin{pmatrix} x_1 & \ldots & x_{\lfloor\frac{n}{2}\rfloor+1} & y_1 & \ldots & y_{\lfloor\frac{n}{2}\rfloor} \\ x_{\lfloor\frac{n}{2}\rfloor+1} & \ldots & x_n & y_{\lfloor\frac{n}{2}\rfloor+2} & \ldots & y_n \\ y_1 & \ldots & y_{\lfloor\frac{n}{2}\rfloor+1} & x_1 & \ldots & x_{\lfloor\frac{n}{2}\rfloor} \\ y_{\lfloor\frac{n}{2}\rfloor+1} & \ldots & y_n & x_{\lfloor\frac{n}{2}\rfloor+2} & \ldots & x_n \end{pmatrix}$

An algorithm for function $T_{2n}$ gives a set of amplitudes $<a_1, a_2, a_3, a_4>$. To assure exactness property of the algorithm, $<a_1, a_2, a_3, a_4>$ should fit condition: one of $a_i$ ($i = 1..4$) is 1 or -1, others are zeros. With regard to Lemma 1, we receive such a distribution of amplitudes, that after applying $U_1$ we get the required above distribution with only one $|a_i| = 1$ and other amplitudes equal to zero. □

**Lemma 1.** After the chain of queries Q we get such distribution of amplitudes $<a_1, a_2, a_3, a_4>$, where each $|a_i| = \frac{1}{2}$ and there is an even number of negative amplitudes in the distribution.

**Proof.**
$T_4$ and $T_6$ fit this condition; let us consider $T_4$ and $T_6$ to be the base of induction.
We assume $T_{2n}$ gives us necessary distribution of amplitudes after the query Q is applied: there appears even number of negative amplitudes.
It is necessary to prove that adding two more variables $x_{n+1}$ and $y_{n+1}$ gives new $T_{2(n+1)}$ exact quantum query algorithm.
If n is even, then adding two more variables is equivalent to substituting query matrix $Q_{even}$ for $Q'_{even}$.

$Q'_{even} = \begin{pmatrix} x_1 & \ldots & x_{\frac{n}{2}+1} & y_1 & \ldots & y_{\frac{n}{2}} \\ x_{\frac{n}{2}+1} & \ldots & x_{n+1} & y_{\frac{n}{2}+2} & \ldots & y_{n+1} \\ y_1 & \ldots & y_{\frac{n}{2}+1} & x_1 & \ldots & x_{\frac{n}{2}} \\ y_{\frac{n}{2}+1} & \ldots & y_{n+1} & x_{\frac{n}{2}+2} & \ldots & x_{n+1} \end{pmatrix}$, $Qdif = \begin{pmatrix} x_{\frac{n}{2}+1} \\ x_{n+1} y_{\frac{n}{2}+1} y_{n+1} \\ y_{\frac{n}{2}+1} \\ y_{n+1} x_{\frac{n}{2}+1} x_{n+1} \end{pmatrix}$

$Q_{dif}$ changes amplitudes after applying queries from $Q_{even}$ in such a way that it becomes equivalent to applying $Q'_{even}$. It can be easily checked that $Q_{dif}$ flips an even number of amplitudes to the opposite, thus maintaining an even number of negative amplitudes for the new query sequence Q'$_{even}$.

If n is odd, then adding two more variables is equivalent to substituting query matrix Q$_{odd}$ for Q'$_{odd}$.

$$Q'_{odd} = \begin{pmatrix} x_1 & \ldots & x_{\lfloor \frac{n}{2} \rfloor+1} & y_1 & \ldots & y_{\lfloor \frac{n}{2} \rfloor+1} \\ x_{\lfloor \frac{n}{2} \rfloor+2} & \ldots & x_{n+1} & y_{\lfloor \frac{n}{2} \rfloor+2} & \ldots & y_{n+1} \\ y_1 & \ldots & y_{\lfloor \frac{n}{2} \rfloor+1} & x_1 & \ldots & x_{\lfloor \frac{n}{2} \rfloor+1} \\ y_{\lfloor \frac{n}{2} \rfloor+2} & \ldots & y_{n+1} & x_{\lfloor \frac{n}{2} \rfloor+2} & \ldots & x_{n+1} \end{pmatrix}, \quad Qdif = \begin{pmatrix} y_{\lfloor \frac{n}{2} \rfloor+1} \\ x_{\lfloor \frac{n}{2} \rfloor+1} x_{n+1} y_{n+1} \\ x_{\lfloor \frac{n}{2} \rfloor+1} \\ y_{\lfloor \frac{n}{2} \rfloor+1} y_{n+1} x_{n+1} \end{pmatrix}$$

Q$_{dif}$, like in the previous case, changes amplitudes after applying queries from Q$_{odd}$ in such a way that it becomes equivalent to applying Q'$_{odd}$. It can be easily checked that Q$_{dif}$ flips an even number of amplitudes to the opposite, thus maintaining an even number of negative amplitudes for the new query sequence Q'$_{odd}$. □

We have proved that exact quantum algorithm computing the function T$_{2n}$ can be changed to compute T$_{2(n+1)}$ by substituting query sequences Q$_{even}$ for Q'$_{even}$ or Q$_{odd}$ for Q'$_{odd}$. Hence, we have an unlimited set of exact quantum algorithms computing corresponding functions T$_{2n}$ with n queries only, while deterministic algorithm can do it with 2n queries(this property is provided by sensitivity of function T$_{2n}$ on input 00….0: flipping one variable value flips the value of function).

# 4 Algorithm Transformation Methods

In this section we introduce quantum query algorithm transformation methods that can be useful for enlarging a set of exactly computable Boolean functions. Each method receives exact quantum query algorithm on input, processes it as defined, and as a result slightly different exact algorithm is obtained that computes another Boolean function.

## 4.1 Output value assignment inversion

The first method is the simplest one. All we need to do with original algorithm is to change assigned function value for each output to the opposite.

| *First Transformation Method - Output value assignment inversion* |
|---|
| **Input.** An arbitrary exact QQA that computes *f(X)*. |
| **Transformation actions.** <br> • For each algorithm output change assigned value of function to opposite. <br>   If original assignment was $QM = (\alpha_1 \equiv k_1,...,\alpha_m \equiv k_m)$, where $k_i \in \{0,1\}$, <br>   Then it is transformed to $QM' = (\alpha_1 \equiv \overline{k}_1,...,\alpha_m \equiv \overline{k}_m)$, where $\overline{k}_i = 1-k_i$. |
| **Output.** An exact QQA that computes $\overline{f}(X)$. |

Box 1 Description of the First Transformation Method.

**Theorem 2.** An arbitrary exact QQA for a function *f* can be transformed to compute $\bar{f}$ without increasing number of queries.

**Proof.** Transformation algorithm is presented in Box 1. We change only output value assignment; number of queries remains the same. The fact that new algorithm will compute $\bar{f}$ is obvious and follows straightforward from the definition of QQA. □

**Corollary 1.** For an arbitrary Boolean function $Q_E(f) = Q_E(\bar{f})$.

**Proof.** It follows straightforward from Theorem 2. Optimal exact QQA for *f* can be transformed to compute $\bar{f}$ without increasing complexity and vice versa. □

**Example of application.** Let us recall, that in section 3.2 we presented exact QQA for the task to determine if two binary strings are equal. Now we can easily convert our algorithm to compute opposite task, i.e. to check if two binary strings are NOT equal. What we have to do is simply to change output evaluation rule from $QM = (\alpha_1 \equiv 1, \alpha_2 \equiv 0, \alpha_3 \equiv 0, \alpha_4 \equiv 0)$ to $QM' = (\alpha_1 \equiv 0, \alpha_2 \equiv 1, \alpha_3 \equiv 1, \alpha_4 \equiv 1)$.

## 4.2 Output value assignment permutation

Describing next method we will limit ourselves to using only exact QQA with specific properties as an input for transformation method. We define the property of exact QQA designed in a model considered in this paper.

**Property 1.** We say that exact QQA satisfies *Property 1* IFF on any input amplitude distribution before final measurement is distinct, that is for exactly one amplitude $\alpha_i$ it holds true that $|\alpha_i|^2 = 1$ (implying that quantum basic state $|i\rangle$ will be observed after a measurement). For all other amplitudes it holds true that $|\alpha_j|^2 = 0$, for $\forall j \neq i$.

**Example 1.**

Distinct distributions: $(1,0,0,0), \left(0,0,\frac{1}{\sqrt{2}}+\frac{1}{\sqrt{2}}i,0\right)$. Not distinct: $\left(\frac{1}{2},\frac{1}{2},\frac{1}{2},\frac{1}{2}\right)\left(\frac{1}{\sqrt{2}},0,0,-\frac{1}{\sqrt{2}}i\right)$.

---

*Second Transformation Method - Output value assignment permutation*

**Input.**
- An exact QQA satisfying *Property 1* that computes *f(X)*.
- Permutation $\sigma$ of the set $OutputValues = \{k_1, k_2, ..., k_m\}$.

**Transformation actions.**
- Permute function values assigned to outputs in order specified by $\sigma$.
  
  If original assignment was $QM = (\alpha_1 \equiv k_1, ..., \alpha_m \equiv k_m)$, where $k_i \in \{0,1\}$,
  
  Then it is transformed to $QM' = (\alpha_1 \equiv \sigma(k_1), ..., \alpha_m \equiv \sigma(k_m))$.

**Output.** An exact QQA for some function *g(X)*.

---

Box 2 Description of the Second Transformation Method.

**Example 2.**

We will explain a method described in the Box 2 with a short example. Let us recall an algorithm described in section 3.1 that computes 3-variable equality function. Original output value assignment there is $QM = (\alpha_1 \equiv 1, \alpha_2 \equiv 0, \alpha_3 \equiv 0, \alpha_4 \equiv 0)$. For this example we will use the following permutation $\sigma = \begin{pmatrix} k_1 & k_2 & k_3 & k_4 \\ k_4 & k_2 & k_3 & k_1 \end{pmatrix}$, what means that transformed output value assignment is:

$QM' = (\alpha_1 \equiv \sigma(k_1), \alpha_2 \equiv \sigma(k_2), \alpha_3 \equiv \sigma(k_3), \alpha_4 \equiv \sigma(k_4)) = (\alpha_1 \equiv k_4, \alpha_2 \equiv k_2, \alpha_3 \equiv k_3, \alpha_4 \equiv k_1) =$
$= (\alpha_1 \equiv 0, \alpha_2 \equiv 0, \alpha_3 \equiv 0, \alpha_4 \equiv 1)$

There exist only two more non-trivial permutations, which gives possible outcomes $QM' = (\alpha_1 \equiv 0, \alpha_2 \equiv 1, \alpha_3 \equiv 0, \alpha_4 \equiv 0)$ and $QM' = (\alpha_1 \equiv 0, \alpha_2 \equiv 0, \alpha_3 \equiv 1, \alpha_4 \equiv 0)$.

**Theorem 3.** Application of the Second Transformation Method doesn't break the exactness of QQA and allows computing different Boolean function without increasing number of queries.

**Proof.**

Let us show that output value permutation method application for exact QQA satisfying *Property 1* allows computing some different Boolean function exactly. The essence of *Property 1* is that before the measurement we always obtain non-zero amplitude value in exactly one output. Since function value is clearly specified for each output we would always observe some specific function value with probability 1 for any input. □

We would like to say few words about the function $g(X)$ that will be computable by new algorithm obtained after application of the Second Transformation Method. The essence and structure of new function strictly depends on internal properties of original algorithm. To understand and explicitly define new function one needs to inspect original algorithm behavior on each input and construct a truth table for new output value assignment.

For instance, using algorithm presented in section 3.1 as a base we obtained algorithms that compute the following Boolean function depending on output value assignment:

| Output value assignment | Boolean function |
|---|---|
| $QM = (\alpha_1 \equiv 1, \alpha_2 \equiv 0, \alpha_3 \equiv 0, \alpha_4 \equiv 0)$ | $F_3^1(X) = \neg(x_1 \oplus x_2) \wedge \neg(x_2 \oplus x_3)$ |
| $QM' = (\alpha_1 \equiv 0, \alpha_2 \equiv 1, \alpha_3 \equiv 0, \alpha_4 \equiv 0)$ | $F_3^2(X) = (x_1 \oplus x_2) \wedge \neg(x_2 \oplus x_3)$ |
| $QM' = (\alpha_1 \equiv 0, \alpha_2 \equiv 0, \alpha_3 \equiv 1, \alpha_4 \equiv 0)$ | $F_3^3(X) = (x_1 \oplus x_2) \wedge (x_2 \oplus x_3)$ |
| $QM' = (\alpha_1 \equiv 0, \alpha_2 \equiv 0, \alpha_3 \equiv 0, \alpha_4 \equiv 1)$ | $F_3^4(X) = \neg(x_1 \oplus x_2) \wedge (x_2 \oplus x_3)$ |

Table 2. Results of the Second Transformation Method application for EQUALITY$_3$.

## 4.3 Query variable permutation

The next transformation method is relatively obvious as well. This time we are starting from opposite direction. Firstly we transform the definition of function, but then correspondingly adjust an algorithm.

We use a trick of function variable permutation to obtain a slightly different function.

Let $\sigma$ be a permutation of the set $\{1, 2, ..., n\}$, where elements correspond to variable numbers.

By saying that function $g(X)$ is obtained by permutation of $f(X)$ variables we mean the following:

$$g(X) = f\left(x_{\sigma(1)}, x_{\sigma(2)}, ..., x_{\sigma(n)}\right)$$

**Example 3.** We recall the function from section 3.2:

$$F_4(X) = \neg((x_1 \oplus x_3) \vee (x_2 \oplus x_4))$$

By applying the permutation $\sigma = \begin{pmatrix} 1 & 2 & 3 & 4 \\ 2 & 4 & 1 & 3 \end{pmatrix}$ we obtain another function:

$$G_4(X) = \neg((x_2 \oplus x_1) \vee (x_4 \oplus x_3)) = \neg((x_1 \oplus x_2) \vee (x_3 \oplus x_4))$$

In our Third Transformation Method we expand the idea of variable permutation to QQA algorithm definition.

---

*Third Transformation Method – Query variable permutation*

**Input.**
- An arbitrary exact QQA that computes $f_n(X)$.
- Variable number permutation $\sigma$ of the set $VarNum = \{0, 1, ..., n\}$.

**Transformation actions.**
- Apply variable number permutation $\sigma$ to all query transformations.
  If original $i$-th query was defined as $QQ_i = (\alpha_1 \equiv k_1, ..., \alpha_m \equiv k_m)$, $k_i \in \{1, ..., n\}$
  Then it is transformed to $QQ_i' = (\alpha_1 \equiv \sigma(k_1), ..., \alpha_m \equiv \sigma(k_m))$.

**Output.** An exact QQA computing a function $g(X) = f\left(x_{\sigma(1)}, x_{\sigma(2)}, ..., x_{\sigma(n)}\right)$.

---

Box 3 Description of the Third Transformation Method.

**Theorem 4.** Let $A$ be an exact QQA that computes $f$. To construct an exact QQA for a function $g(X) = f\left(x_{\sigma(1)}, x_{\sigma(2)}, ..., x_{\sigma(n)}\right)$ it is enough to apply $\sigma$ to variable numbers in each query of algorithm $A$.

**Proof.** Obvious. If we will apply transformation method described in Box 3, then variable values will influence new algorithm flow according to the order specified by permutation $\sigma$, thus algorithm will compute $g(X)$ instead of $f(X)$. □

## 5 Transformation methods application results

In this section we will demonstrate a set of exactly computable Boolean functions that was obtained by applying transformation methods presented in previous section to the base exact QQA algorithms from section 3.

### 5.1 3-variable functions

Let us recall that in section 3.1 we presented exact QQA for the EQUALITY$_3$ function. Let's see how transformation methods can help us to enlarge a set of exactly computable functions based on existing algorithm.

It comes out that using the First and the Second Transformation Methods we are able to get a set of 3-variable Boolean functions $S3$, where $|S3| = 8$ and for each function classical deterministic complexity is 3, but transformed quantum algorithm uses only 2 queries. Unfortunately the Third Transformation Method doesn't generate any new Boolean function in concerned case of 3-variable functions.

| X | EQUALITY | Output value permutation (2$^{nd}$ method) | | | Output value inversion (1$^{st}$ method) | | | |
|---|---|---|---|---|---|---|---|---|
| | | (0,1,0,0) | (0,0,1,0) | (0,0,0,1) | (0,1,1,1) | (1,0,1,1) | (1,1,0,1) | (1,1,1,0) |
| 000 | 1 | 0 | 0 | 0 | 0 | 1 | 1 | 1 |
| 001 | 0 | 0 | 0 | 1 | 1 | 1 | 1 | 0 |
| 010 | 0 | 0 | 1 | 0 | 1 | 1 | 0 | 1 |
| 011 | 0 | 1 | 0 | 0 | 1 | 0 | 1 | 1 |
| 100 | 0 | 1 | 0 | 0 | 1 | 0 | 1 | 1 |
| 101 | 0 | 0 | 1 | 0 | 1 | 1 | 0 | 1 |
| 110 | 0 | 0 | 0 | 1 | 1 | 1 | 1 | 0 |
| 111 | 1 | 0 | 0 | 0 | 0 | 1 | 1 | 1 |
| $D(f)$ | 3 | 3 | 3 | 3 | 3 | 3 | 3 | 2 |
| $Q_E(f)$ | **2** | **2** | **2** | **2** | **2** | **2** | **2** | **2** |

Table 3. Results of transformation methods application for Algorithm 1 (set S3).

## 5.2   4-variable functions

To construct a set of efficiently computable 4-variable functions we used algorithm for equality of strings of length 2 from section 3.2 as a base for transformation methods. This time all transformation methods gave positive results, 12 functions are presented in Table 4 plus 12 inversed functions for each. Totally we have a set S4, where $|S4| = 24$.

| X | String Equal. (original) | Output value permutation (2$^{nd}$ method) | | | $\sigma_{VarNum} = \begin{pmatrix}1234\\1324\end{pmatrix}$ 3$^{rd}$ + 2$^{nd}$ method | | | | $\sigma_{VarNum} = \begin{pmatrix}1234\\3124\end{pmatrix}$ 3$^{rd}$ + 2$^{nd}$ method | | | |
|---|---|---|---|---|---|---|---|---|---|---|---|---|
| | | $\begin{pmatrix}0\\1\\0\\0\end{pmatrix}$ | $\begin{pmatrix}0\\0\\1\\0\end{pmatrix}$ | $\begin{pmatrix}0\\0\\0\\1\end{pmatrix}$ | $\begin{pmatrix}1\\0\\0\\0\end{pmatrix}$ | $\begin{pmatrix}0\\1\\0\\0\end{pmatrix}$ | $\begin{pmatrix}0\\0\\1\\0\end{pmatrix}$ | $\begin{pmatrix}0\\0\\0\\1\end{pmatrix}$ | $\begin{pmatrix}1\\0\\0\\0\end{pmatrix}$ | $\begin{pmatrix}0\\1\\0\\0\end{pmatrix}$ | $\begin{pmatrix}0\\0\\1\\0\end{pmatrix}$ | $\begin{pmatrix}0\\0\\0\\1\end{pmatrix}$ |
| 0000 | **1** | 0 | 0 | 0 | 0 | 0 | 0 | 1 | 1 | 0 | 0 | 0 |
| 0001 | 0 | 1 | 0 | 0 | 0 | 1 | 0 | 0 | 0 | 1 | 0 | 0 |
| 0010 | 0 | 0 | 1 | 0 | 0 | 1 | 0 | 0 | 0 | 0 | 1 | 0 |
| 0011 | 0 | 0 | 0 | 1 | 0 | 0 | 0 | 1 | 0 | 0 | 0 | 1 |
| 0100 | 0 | 1 | 0 | 0 | 0 | 0 | 1 | 0 | 0 | 0 | 1 | 0 |
| 0101 | **1** | 0 | 0 | 0 | 1 | 0 | 0 | 0 | 0 | 0 | 0 | 1 |
| 0110 | 0 | 0 | 0 | 1 | 1 | 0 | 0 | 0 | 1 | 0 | 0 | 0 |
| 0111 | 0 | 0 | 1 | 0 | 0 | 0 | 1 | 0 | 0 | 1 | 0 | 0 |
| 1000 | 0 | 0 | 1 | 0 | 0 | 0 | 1 | 0 | 0 | 1 | 0 | 0 |
| 1001 | 0 | 0 | 0 | 1 | 1 | 0 | 0 | 0 | 1 | 0 | 0 | 0 |
| 1010 | **1** | 0 | 0 | 0 | 1 | 0 | 0 | 0 | 0 | 0 | 0 | 1 |
| 1011 | 0 | 1 | 0 | 0 | 0 | 0 | 1 | 0 | 0 | 0 | 1 | 0 |
| 1100 | 0 | 0 | 0 | 1 | 0 | 0 | 0 | 1 | 0 | 0 | 0 | 1 |
| 1101 | 0 | 0 | 1 | 0 | 0 | 1 | 0 | 0 | 0 | 0 | 1 | 0 |
| 1110 | 0 | 1 | 0 | 0 | 0 | 1 | 0 | 0 | 0 | 1 | 0 | 0 |
| 1111 | **1** | 0 | 0 | 0 | 0 | 0 | 0 | 1 | 1 | 0 | 0 | 0 |
| $D(f)$ | 4 | 4 | 4 | 4 | 4 | 4 | 4 | 4 | 4 | 4 | 4 | 4 |
| $Q_E(f)$ | **2** | **2** | **2** | **2** | **2** | **2** | **2** | **2** | **2** | **2** | **2** | **2** |

Table 4. Results of transformation methods application for Algorithm 2 (set S4).

# 6 Algorithm Designing Methods

In this section we will present several quantum query algorithm designing methods. Each method requires explicitly specified exact QQA algorithms on input, but as a result a bounded-error QQA for more complex function is constructed. Our methods maintain quantum query complexity for complex function in comparison to increased deterministic complexity, thus enlarging the gap between classical and quantum complexities of an algorithm.

## 6.1 Obtaining a gap $D(f)=6$ vs. $Q_{3/4}(f)=2$

In section 3.1 we presented exact QQA for the function $EQUALITY_3(X) = \neg(x_1 \oplus x_2) \wedge \neg(x_2 \oplus x_3)$. Now we will try to solve a bit more complex problem. We consider composite Boolean function, where two instances of $EQUALITY_3$ are joined with logical AND operation:

$$EQUALITY_3^{\wedge 2}(x_1, x_2, x_3, x_4, x_5, x_6) = EQUALITY_3(x_1, x_2, x_3) \wedge EQUALITY_3(x_4, x_5, x_6)$$

$$EQUALITY_3^{\wedge 2}(x_1, x_2, x_3, x_4, x_5, x_6) = (\neg(x_1 \oplus x_2) \wedge \neg(x_2 \oplus x_3)) \wedge (\neg(x_4 \oplus x_5) \wedge \neg(x_5 \oplus x_6))$$

In other words, we applied a pattern defined as $EQUALITY_3$ to the first half of variables, then the same pattern to the second half of variables, finally we joined both terms with AND operation. We denote obtained function $EQUALITY_3^{\wedge 2}$ to designate the fact that two blocks defined by $EQUALITY_3$ were joined using $\wedge$ operation.

To evaluate deterministic complexity for that and further functions we will use the following lemma.

**Lemma 2.** We consider a Boolean function expressed in a form $F(X) = f_1(X_1) \wedge ... \wedge f_n(X_n)$, i.e. several variable blocks constrained by a pattern of functions $f_1,...,f_n$ are joined with AND operation. We denote by $s_1(f)$ sensitivity of $f$ on accepting input. Then the following statement is true:  If $\forall f_i : s_1(f_i) = n_i$, where $n_i$ - number of variables,

Then $s_1(F) = s_1(f_1) + ... + s_1(f_n) = N$, where $N$ is number of variables of $F$.

**Proof.**
Let's consider some accepting input for $F$: $X = X_1 X_2 ... X_n$, where $X_i$ is accepting input for $f_i$ such that $s_1(X_i) = n_i$. We have $(\forall i : f_i(X_i) = 1) \Rightarrow (F(X) = 1)$. From $s_1(X_i) = n_i$ follows that change of any input bit in $X_i$ will inverse value of $f_i$. If we change $j$-th bit of $X_i$ then we have $(f_i(X_i^{(j)}) = 0) \Rightarrow (F(X) = 0)$, because of the essence of AND joining operation. So, we are not allowed to flip any input bit without changing value of function $F$, thus $s_1(F) = N$. □

**Deterministic complexity.** $D(EQUALITY_3^{\wedge 2}) = 6$

**Proof.** We use Lemma 2 to prove this estimation. Let us recall the definition of our function:
$$EQUALITY_3^{\wedge 2}(X) = EQUALITY_3(X_1) \wedge EQUALITY_3(X_2)$$

For sub-function we have $s_1(EQUALITY_3) = 3$ (check either $X=000$ or $X=111$). Thus by Lemma 2 we have $s_1(EQUALITY_3^{\wedge 2}) = s_1(EQUALITY) + s_1(EQUALITY) = 6$. From the fact that $D(f) \geq s(f)$ follows $D(EQUALITY_3^{\wedge 2}) = 6$. □

Our approach in designing an algorithm for $EQUALITY_3^{\wedge 2}$ is to employ quantum parallelism and superposition principle. We extend quantum system with additional qubits, now there will be three qubits instead of two qubits used by original algorithm. We execute algorithm pattern defined by original algorithm for $EQUALITY_3$ in parallel for both blocks of variables of $EQUALITY_3^{\wedge 2}$. Finally we apply additional quantum gate to correlate amplitude distribution. Algorithm flow is depicted explicitly in Figure 5.

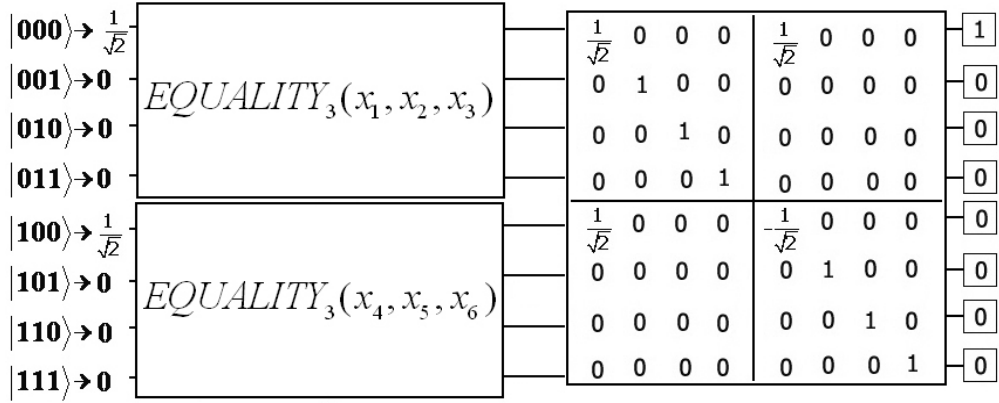

Fig. 5 Bounded-error QQA for $EQUALITY_3^{\wedge 2}$

**Quantum complexity.** Quantum query algorithm presented in Figure 5 computes $EQUALITY_3^{\wedge 2}$ using 2 queries with correct answer probability $p = \frac{3}{4}$: $Q_{3/4}(EQUALITY_3^{\wedge 2}) = 2$.

**Proof.**
To calculate probabilities of obtaining correct function value it is enough to examine 4 cases depending on the value of each term of $EQUALITY_3^{\wedge 2}$. Results are presented in Table 5. We use wildcards "?" and "*" to denote that exactly one value under the same wildcard is $\pm \frac{1}{\sqrt{2}}$ (we don't care which one), but all others are zeroes.

| $EQUALITY_3$ $(x_1, x_2, x_3)$ | $EQUALITY_3$ $(x_4, x_5, x_6)$ | Amplitude distribution before last gate | Amplitude distribution after last gate | $p("1")$ |
|---|---|---|---|---|
| 0 | 0 | $(0,?,?,?,0,*,*,*)$ | $(0,?,?,?,0,*,*,*)$ | **0** |
| 0 | 1 | $\left(0,?,?,?,\frac{1}{\sqrt{2}},0,0,0\right)$ | $\left(\frac{1}{2},?,?,?,-\frac{1}{2},0,0,0\right)$ | **1/4** |
| 1 | 0 | $\left(\frac{1}{\sqrt{2}},0,0,0,?,?,?\right)$ | $\left(\frac{1}{2},0,0,0,\frac{1}{2},?,?,?\right)$ | **1/4** |
| 1 | 1 | $\left(\frac{1}{\sqrt{2}},0,0,0,\frac{1}{\sqrt{2}},0,0,0\right)$ | $(1,0,0,0,0,0,0,0)$ | **1** |

Table 5 Calculation of probabilities depending on algorithm flow for $EQUALITY_3^{\wedge 2}$.

So, we have $p("1") = 1$ and $p("0") = 3/4$, we did not use additional queries, thus estimation $Q_{3/4}(EQUALITY_3^{\wedge 2}) = 2$ is proved. □

## 6.2 First Designing Method

In this section we will generalize approach used in previous section and introduce our first algorithm designing method. To be able to use generalized version of approach we will limit ourselves to examining only exact QQA with specific properties as an input base for designing method.

**Property 2+** We say that exact QQA satisfies *Property2+* IFF there is exactly one accepting basic state and on any input for its amplitude $\alpha \in C^n$ only two values are possible before the final measurement: either $\alpha = 0$ or $\alpha = 1$.

The essence of *Property2+* is that in the case of executing QQA on accepting input we will always get clear "+1" in that single output with assigned function value 1. We define similar property for the case when clear "-1" is obtained in accepting amplitude before measurement.

**Property 2-** We say that exact QQA satisfies *Property2-* IFF there is exactly one accepting basic state and on any input for its amplitude $\alpha \in C^n$ only two values are possible before the final measurement: either $\alpha = 0$ or $\alpha = -1$.

In the process of method application we will need the following lemma regarding algorithms satisfying described properties.

**Lemma 3.** It is possible to transform algorithm that satisfies *Property2-* to algorithm satisfying *Property2+* by applying additional unitary transformation.

**Proof.** Let's assume that we have QQA satisfying *Property2-* and $k$ is the number of accepting output. To transform algorithm to satisfy *Property2+* apply the following quantum gate $U$, which differs from identity matrix in only one element:

$$U = u_{ij} = \begin{cases} 0, & \text{if } i \neq j \\ 1, & \text{if } i = j \neq k \\ -1, & \text{if } i = j = k \end{cases}$$

After additional gate all amplitude values in case of accepting input will be converted from "-1" to "+1", thus algorithm will satisfy *Property2+*. □

---

*First Designing Method*

**Input.**

- Two exact QQAs A1 and A2 satisfying *Property2+* or *Property2-* that compute correspondingly Boolean functions $f_1(X_1)$ and $f_2(X_2)$

**Transformation actions.**

1. If A1 or A2 satisfy *Property2-* then transform it to satisfy *Property2+* as described in the proof of Lemma 3.
2. If A1 and A2 utilize quantum systems of different size (different number of qubits are used), then extend the smallest one with auxiliary space to obtain equal number of amplitudes. We denote the dimension of obtained Hilbert spaces with $m$, so each quantum system consist of $\log(m)$ qubits.
3. For new algorithm utilize a quantum system with $2m$ amplitudes.
   $\log(2m) = \log(m)+1$, so exactly one additional qubit is required.
4. Combine unitary transformations and queries of A1 and A2 in the following way:

$$U_i = \begin{pmatrix} U_i^1 & O \\ O & U_i^2 \end{pmatrix},$$ here $O$'s are $m \times m$ zero-matrices, $U_i^1$ and $U_i^2$ are either unitary transformations or query transformations of $A1$ and $A2$.

5. Start computation from the state $\langle \psi | = \left( \frac{1}{\sqrt{2}}, 0, ..., 0, \frac{1}{\sqrt{2}}, 0, ..., 0 \right)$, here $\frac{1}{\sqrt{2}}$ is present in first and $(m+1)$-th positions of state vector.

6. Before the final measurement apply additional unitary gate. Let's denote the positions of accepting outputs of A1 and A2 by $acc_1$ and $acc_2$. Then the final gate is defined as follows:

$$U = u_{ij} = \begin{cases} 1, & \text{if } (i = j) \& (i \neq acc_1) \& (i \neq (m + acc_2)) \\ \frac{1}{\sqrt{2}}, & \text{if } (i = j = acc_1) \\ \frac{1}{\sqrt{2}}, & \text{if } (i = acc_1) \& (j = (m + acc_2)) \\ \frac{1}{\sqrt{2}}, & \text{if } (i = (m + acc_2)) \& (j = acc_1) \\ -\frac{1}{\sqrt{2}}, & \text{if } (i = j = (m + acc_2)) \\ 0, & \text{otherwise} \end{cases}$$

7. Define as accepting output exactly one basic state $|acc_1\rangle$ that corresponds to accepting output of A1 part of algorithm.

**Output.** A bounded-error QQA $A$ computing a function $F(X) = f_1(X_1) \wedge f_2(X_2)$ with correct answer probability $p = \frac{3}{4}$ and complexity $Q_{3/4}(A) = \max(Q_E(A_1), Q_E(A_2))$.

Box 4. Description of the First Designing Method.

Our last note is that any (non-inversed) algorithm from the set S3 described in section 5.1 can be used as a base for described First Designing Method. So, totally we are able to construct $4*4=16$ different algorithms computing corresponding Boolean functions obtaining a gap $D(f)=6$ vs. $Q_{3/4}(f)=2$ for each function. Evaluation $D(f)=6$ can be obtained for each case using Lemma 2.

## 6.3 Obtaining a gap $D(f)=8$ vs. $Q_{3/4}(f)=2$

Input restrictions of the First Designing Method do not provide enough freedom for constructing bounded-error QQAs ad-hoc. Our aim is to minimize restrictions on input exact QQAs and as a result to extend a set of efficiently computable functions.

Considering algorithms presented in this paper for now we are not able to build bounded-error QQAs for complex functions composed from algorithms for functions from set S4 (section 5.2), because all they satisfy Property1 that is weaker than required Property2$x$.

Now we will try to apply approach described in the First Designing Method to a function from S4. We will try with $STRING\_EQ_4(X) = \neg((x_1 \oplus x_3) \vee (x_2 \oplus x_4))$ discussed in section 3.2.

We apply procedure described in the First Designing Method to Algorithm2 ignoring the fact that it doesn't satisfy required property. The resulting algorithm *Algorithm4* has exactly the same structure as one presented in Figure 5, the difference is that now we execute exact QQA for STRING_EQ$_4$ in parallel instead of EQUALITY$_3$.

We claim that designed algorithm will compute some specific Boolean function and obtained complexity gap is $D(New\_Function) = 8$ vs. $Q_{3/4}(Algorithm4) = 2$.

Let's try to understand the behavior of *Algorithm4* that is influenced by violation of Property2x of sub-algorithms. First let us define next property of exact QQA that extends previous Property2x.

**Property 3** We say that exact QQA satisfies *Property3* IFF there is exactly one accepting basic state and after processing any input its amplitude before the measurement is $\alpha \in \{-1, 0, 1\}$.

Each (non-inversed) algorithm from the set S4, including that one for STRING_EQ$_4$, satisfies Property3. We would like to note, that defined properties form the following exact QQA hierarchy:

$$\text{Property2x} \subseteq \text{Property3} \subseteq \text{Property1}.$$

For *Algorithm4* only the following amplitudes of accepting output are possible before the final measurement: $\alpha \in \{0, \pm\frac{1}{2}, \pm 1\}$. Details are presented in a table below.

| $STRING\_EQ_4$ $(x_1, x_2, x_3, x_4)$ | $STRING\_EQ_4$ $(x_5, x_6, x_7, x_8)$ | Amplitude distribution before last gate | Amplitude of accepting output before measurement | $p("1")$ |
|---|---|---|---|---|
| 0 | 0 | $(0,?,?,?,0,*,*,*)$ | 0 | **0** |
| 0 | 1 | $\left(0,?,?,?,\pm\frac{1}{\sqrt{2}},0,0,0\right)$ | $\pm\frac{1}{2}$ | **1/4** |
| 1 | 0 | $\left(\pm\frac{1}{\sqrt{2}},0,0,0,0,?,?,?\right)$ | $\pm\frac{1}{2}$ | **1/4** |
| 1 | 1 | $\pm\left(\frac{1}{\sqrt{2}},0,0,0,-\frac{1}{\sqrt{2}},0,0,0\right)$ | 0 | **0** |
| 1 | 1 | $\pm\left(\frac{1}{\sqrt{2}},0,0,0,\frac{1}{\sqrt{2}},0,0,0\right)$ | $\pm 1$ | **1** |

Table 6. Calculation of probabilities for Algorithm4.

Now we see that *Algorithm4* does NOT compute the following function:

$STRING\_EQ_4^{\wedge 2}(X) = STRING\_EQ_4(x_1, x_2, x_3, x_4) \wedge STRING\_EQ_4(x_5, x_6, x_7, x_8)$

as it was in the case of the First Designing Method with algorithms satisfying Property2x.

When examining behavior of Algorithm2 for STRING_EQ$_4$ we see that there are totally 4 accepting inputs: $X_{ACCEPT} = \{0000, 0101, 1010, 1111\}$. One half of them {0000, 1111} gives amplitude value "+1" in accepting output before measurement, but other half {0101, 1010} gives "-1". From the last row of Table 6 we conclude that new function value is 1 iff both sub-algorithms executed in parallel produced equal amplitudes in corresponding accepting states before the last gate. Thus final conclusion is formulated in the following corollary.

**Corollary 2.** Bounded-error QQA *Algorithm4* is computing Boolean function defined as:

$$STRING\_EQ_2^{\|}(X) = \begin{cases} 1, & \text{if } X \in \{00000000,\ 00001111,\ 11110000,\ 11111111\} \\ 1, & \text{if } X \in \{01010101,\ 01011010,\ 10101010,\ 10101010\} \\ 0, & \text{otherwise} \end{cases}$$

and complexity is $Q_{3/4}(\text{Algorithm4}) = 2$.

Deterministic complexity similarly to previous examples can be evaluated by checking function sensitivity on any accepting input, thus $D(STRING\_EQ_2^{\|}) = 8$.

## 6.4 Second Designing Method

In this section we will sum up results obtained in previous section and will formulate it as the Second Designing Method.

To simplify a way to define Boolean functions in this section we introduce the following notation. We denote a set of accepting inputs for Boolean function *F* by $Acc_F$. While discussing exact QQA satisfying Property3 we define also the following sets:

$$Acc_F^+ = \{X \in Acc_F \mid \text{accepting output amplitude before measurement is } +1\}$$

$$Acc_F^- = \{X \in Acc_F \mid \text{accepting output amplitude before measurement is -1}\}$$

For algorithms satisfying Property3 it holds true that $Acc_F = Acc_F^+ \cup Acc_F^-$.

---

***Second Designing Method***

**Input.**
- Two exact QQAs A1 and A2 satisfying *Property3* that compute correspondingly Boolean functions $f_1(X_1)$ and $f_2(X_2)$

**Transformation actions.**
- Perform steps 2-7 described in the First Designing Method (Box 4).

**Output.** A bounded-error QQA *A* computing a function $F(X)$ defined below with correct answer probability $p = \dfrac{3}{4}$ and complexity $Q_{3/4}(A) = \max(Q_E(A_1), Q_E(A_2))$.

$$F(X) = \begin{cases} 1, & \text{if } X \in (Acc_{f1}^+ \times Acc_{f2}^+) \cup (Acc_{f1}^- \times Acc_{f2}^-) \\ 0, & \text{otherwise} \end{cases}$$

---

Box 5. Description of the Second Designing Method.

## 6.5 Obtaining a gap $D(f)=9$ vs. $Q_{9/16}(f)=2$

Now let us try to increase the effect gained by employing quantum parallelism. In previous examples we executed already designed exact QQA in two parallel threads. Let's see what happens in case of four parallel threads. We will take as a pattern well discussed function EQUALITY$_3$ from section 3.1. The main idea is to execute 4 instances of algorithm in parallel, adjusting all algorithm parameters in appropriate way.

Starting amplitude distribution will be: $\langle \vec{0} | = \left( \frac{1}{2}, 0, 0, 0, \frac{1}{2}, 0, 0, 0, \frac{1}{2}, 0, 0, 0, \frac{1}{2}, 0, 0, 0 \right)$. Unitary transformations and queries will be 16x16 matrices obtained in a way $U = I \otimes U_{orig}$. The only accepting basic state will be $|0000\rangle$ that corresponds to the first amplitude. Finally, we will apply specific gates $U'$ and $U''$ to correlate amplitude distribution in appropriate way (empty matrix cells here correspond to "0").

$$U' = \begin{pmatrix} \frac{1}{\sqrt{2}} & 0 & 0 & 0 & \frac{1}{\sqrt{2}} & 0 & 0 & 0 & 0 & 0 & 0 & 0 & 0 & 0 & 0 & 0 \\ 0 & 1 & & & 0 & & & & 0 & & & & 0 & & & 0 \\ 0 & & 1 & & 0 & & & & 0 & & & & 0 & & & 0 \\ 0 & & & 1 & 0 & & & & 0 & & & & 0 & & & 0 \\ \frac{1}{\sqrt{2}} & 0 & 0 & 0 & \frac{-1}{\sqrt{2}} & & & & 0 & & & & 0 & & & 0 \\ 0 & & & & & 1 & & & 0 & & & & 0 & & & 0 \\ 0 & & & & & & 1 & & 0 & & & & 0 & & & 0 \\ 0 & & & & & & & 1 & 0 & & & & 0 & & & 0 \\ 0 & 0 & 0 & 0 & 0 & 0 & 0 & 0 & \frac{1}{\sqrt{2}} & 0 & 0 & 0 & \frac{1}{\sqrt{2}} & 0 & 0 & 0 \\ 0 & & & & & & & & 0 & 1 & & & 0 & & & 0 \\ 0 & & & & & & & & 0 & & 1 & & 0 & & & 0 \\ 0 & & & & & & & & 0 & & & 1 & 0 & & & 0 \\ 0 & & & & & & & & \frac{1}{\sqrt{2}} & 0 & 0 & 0 & \frac{-1}{\sqrt{2}} & & & 0 \\ 0 & & & & & & & & 0 & & & & & 1 & & 0 \\ 0 & & & & & & & & 0 & & & & & & 1 & 0 \\ 0 & 0 & 0 & 0 & 0 & 0 & 0 & 0 & 0 & 0 & 0 & 0 & 0 & 0 & 0 & 1 \end{pmatrix} \quad U'' = \begin{pmatrix} \frac{1}{\sqrt{2}} & 0 & 0 & 0 & 0 & 0 & 0 & 0 & \frac{1}{\sqrt{2}} & 0 & 0 & 0 & 0 & 0 & 0 & 0 \\ 0 & 1 & & & & & & & 0 & & & & & & & 0 \\ 0 & & 1 & & & & & & 0 & & & & & & & 0 \\ 0 & & & 1 & & & & & 0 & & & & & & & 0 \\ 0 & & & & 1 & & & & 0 & & & & & & & 0 \\ 0 & & & & & 1 & & & 0 & & & & & & & 0 \\ 0 & & & & & & 1 & & 0 & & & & & & & 0 \\ 0 & & & & & & & 1 & 0 & & & & & & & 0 \\ \frac{1}{\sqrt{2}} & 0 & 0 & 0 & 0 & 0 & 0 & 0 & \frac{-1}{\sqrt{2}} & 0 & 0 & 0 & 0 & 0 & 0 & 0 \\ 0 & & & & & & & & 0 & 1 & & & & & & 0 \\ 0 & & & & & & & & 0 & & 1 & & & & & 0 \\ 0 & & & & & & & & 0 & & & 1 & & & & 0 \\ 0 & & & & & & & & 0 & & & & 1 & & & 0 \\ 0 & & & & & & & & 0 & & & & & 1 & & 0 \\ 0 & & & & & & & & 0 & & & & & & 1 & 0 \\ 0 & 0 & 0 & 0 & 0 & 0 & 0 & 0 & 0 & 0 & 0 & 0 & 0 & 0 & 0 & 1 \end{pmatrix}$$

Derived quantum algorithm *Algorithm5* is presented in Figure 6.

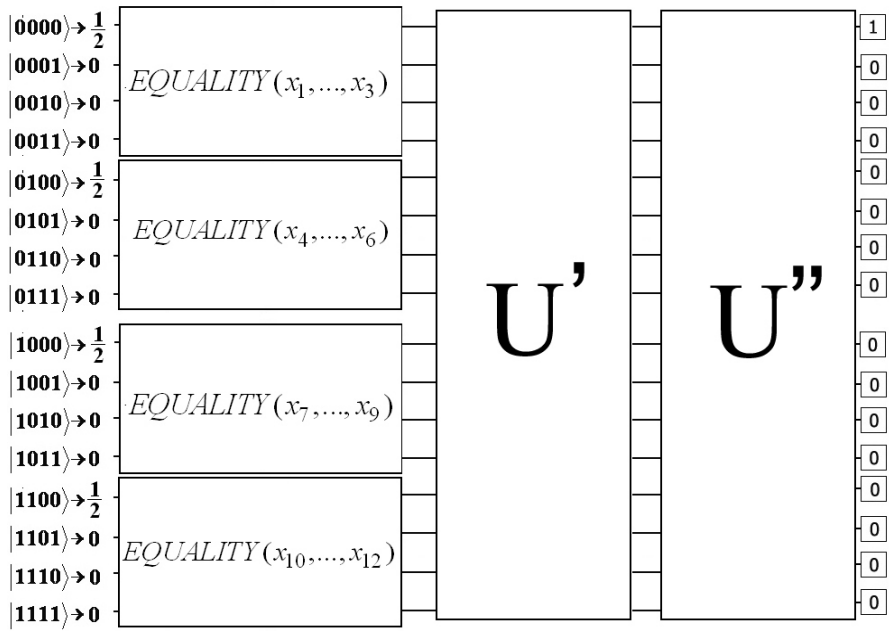

Fig. 6 Quantum query algorithm Algorithm 5.

After examination of algorithm computational flow and calculation of probabilities we obtained result that is formulated in the next corollary.

**Corollary 3.** Bounded-error QQA *Algorithm5* is computing Boolean function defined as:

$$F(x_1,...,x_{12})=1 \Leftrightarrow \begin{pmatrix} \text{Not less than 3 functions from:} \\ EQUALITY(x_1,...,x_3), EQUALITY(x_4,...,x_6) \\ EQUALITY(x_7,...,x_9), EQUALITY(x_{10},...,x_{12}) \\ \text{give value "1".} \end{pmatrix}$$

and complexity is $Q_{9/16}(\text{Algorithm5}) = 2$.

Comparing to deterministic complexity we did not achieve maximal possible gap this time. From the definition of function $F$ we find that sensitivity is $s(F) = 9$, thus in this case we can only register a gap $D(f)=9$ vs. $Q_{9/16}(f)=2$.

## 6.6 Third Designing Method

Our Third Designing Method is generalization of approach demonstrated in previous section. We leave detailed description of transformation actions to an interested reader as an exercise.

---

*Third Designing Method*

**Input.**
- Four exact QQAs A1, A2, A3, A4 satisfying either *Property2+* or *Property2-* that compute correspondingly Boolean functions $f_1(X_1), f_2(X_2), f_3(X_3), f_4(X_4)$.

**Transformation actions.**
- Combine approach described in previous section with the First transformation method and adjust according to the structure of input exact QQAs.

**Output.** A bounded-error QQA *A* computing a function $F(X)$ defined below with correct answer probability $p = \frac{9}{16}$ and complexity:

$Q_{9/16}(A) = \max(Q_E(A_1), Q_E(A_2), Q_E(A_3), Q_E(A_4))$.

$$F(X)=1 \Leftrightarrow \begin{pmatrix} \text{Not less than 3 functions from:} \\ F_1(X_1), F_2(X_2), F_3(X_3), F_4(X_4) \\ \text{give value "1".} \end{pmatrix}$$

---

Box 6. Description of the Third Designing Method.

We would like to note, that it is technically possible to apply approach described in the Third Designing Method to exact QQAs satisfying Property 3. Definition of computable function will be even more complex, but the most important is that in such a way we can design a lot of different algorithms without increasing number of queries.

# 7 Algorithms for Functions with D(*f*)=2*n* vs. $Q_{3/4}(f)=\left\lceil \frac{n}{2} \right\rceil$

In the end we would like to publish another one result obtained regarding constructing bounded-error QQAs for Boolean functions with non-fixed number of variables defined in general form.

This time number of queries required by quantum algorithm is growing with number of variables, but polynomial complexity gap comparing to deterministic complexity remains the same.

Let us describe this algorithm in the way we did in section 3.3. :

$$|0\rangle \to U_0 \to Q \to U_1 \to [QM],$$

Generalizing Q for $T_{2n}$ :

for even number $n$ $Q_{even} = \begin{pmatrix} x_1 & \ldots & x_{\frac{n}{2}} \\ x_{\frac{n}{2}+1} & \ldots & x_n \\ y_1 & \ldots & y_{\frac{n}{2}} \\ y_{\frac{n}{2}+1} & \ldots & y_n \end{pmatrix}$, for odd $n$ $Q_{odd} = \begin{pmatrix} x_1 & \ldots & x_{\lceil\frac{n}{2}\rceil} \\ x_{\lceil\frac{n}{2}\rceil} & \ldots & x_n \\ y_1 & \ldots & y_{\lceil\frac{n}{2}\rceil} \\ y_{\lceil\frac{n}{2}\rceil} & \ldots & y_n \end{pmatrix}$

this kind of queries is not preserving the property of maintaining even number of negative amplitudes after the chain of queries Q, therefore an algorithm for given *n* computes corresponding function as follows:

1 is returned with probability 1, 0 is returned with probability 3/4. Adding two variables preserves these probabilities of the outcome.

**Problem 4 :** For example, let us define a function $T_6$ (x1,…,x3, y1,…,y3):

| XY | $T_6$ | XY | $T_6$ |
|---|---|---|---|
| 000000 | 1 | 101010 | 1 |
| 000111 | 1 | 101101 | 1 |
| 010010 | 1 | 111000 | 1 |
| 010101 | 1 | 111111 | 1 |
| otherwise | 0 | | |

**Deterministic complexity:** $D(T_6) = 6$. Provided by sensitivity on any accepting input string.

**Algorithm 6.** Quantum algorithm that computes $T_6$(XY) with 2 queries and probability no less than ¾ is presented in Figure 7.

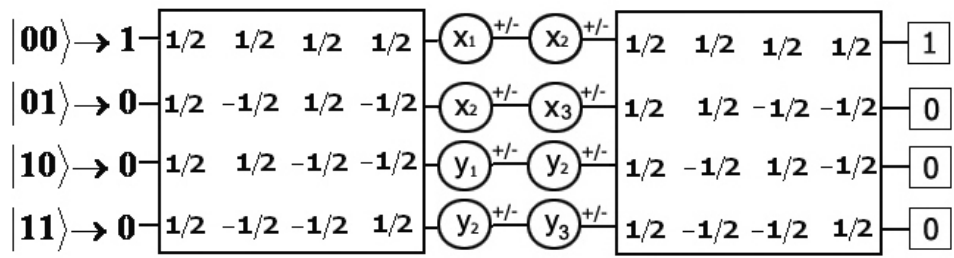

Fig. 7 Exact quantum query algorithm for $T_6$.

Queries can be shown as query matrix $Q = \begin{pmatrix} x1 & x2 \\ x2 & x3 \\ y1 & y2 \\ y2 & y3 \end{pmatrix}$.

**Theorem 5.** For Boolean function $T_{2n}$ of $2n$ variables, where $n \in N$, there exists a bounded-error quantum query algorithm, which computes the function with $\left\lceil \dfrac{n}{2} \right\rceil$ queries, and probability of getting "1" is 1, probability of getting "0" is 3/4.

**Proof.** Similar to the proof of Theorem 1 of section 3.3.

# 8 Conclusion

In this work we describe original and non-trivial exact quantum algorithms for several problems, moreover, all our exact algorithms have the largest possible gap between quantum and deterministic complexities known for today.

We propose techniques that allow transformation of an existing quantum query algorithm for a certain Boolean function so that the resulting algorithm computes a function with other logical structure. We illustrate effect of described methods using previously presented exact algorithms.

Finally we suggest approaches that allow building bounded-error quantum query algorithms based on already known exact algorithms. Our methods do not increase number of queries, but allow computing more complex composite functions of different structures. For example, we are able to design algorithm for a Boolean function presented in conjunctive or disjunctive normal form limited to two terms, using exact algorithms for sub-functions.

Further work in that direction could be to invent new efficient quantum algorithms that exceed already known separation from classical algorithms. Another important direction is improvement of general algorithm designing techniques, which for now still do not provide wishful freedom for the task of constructing quantum algorithms for arbitrary functions.

# 9 References

[1] H. Buhrman and R. de Wolf. "Complexity Measures and Decision Tree Complexity: A Survey". Theoretical Computer Science, v. 288(1): 21-43 (2002).

[2] R. de Wolf. "Quantum Computing and Communication Complexity" University of Amsterdam, 2001.

[3] R.Cleve, A.Ekert, C.Macchiavello, u.c.. *Quantum Algorithms Revisited.* Proceedings of the Royal Society, London, A454, 1998.

[4] J. Gruska. "Quantum Computing" McGraw-Hill, 1999.

[5] M. Nielsen, I. Chuang. "Quantum Computation and Quantum Information" Cambridge University Press 2000. ISBN: 0-521-63503-9.

[6] A.Ambainis. "Quantum query algorithms and lower bounds (survey article)" Proceedings of FOTFS III, to appear.